\documentclass[fleqn,aps,pra,notitlepage,floatfix,longbibliography,twocolumn]{revtex4-2}
\usepackage{amsmath}
\usepackage{graphicx}
\usepackage{hyperref}
\usepackage{bm}
\usepackage{siunitx}

\newcommand{\ket}[1]{|#1\rangle}

\newcommand{\lehigh}{Department of Physics, Lehigh University, Bethlehem, Pennsylvania 18015, USA}

\begin{document}
\title{A self-aligning recirculated crossed  optical dipole trap for lithium atoms}
\author{Ming Lian}
\author{Maximillian Mrozek-McCourt}
\author{Christopher K. Angyal}
\author{Dadbeh Shaddel}
\author{Zachary J. Blogg}
\author{John R. Griffin}
\altaffiliation{Present address: AO Sense, Inc., 
Fremont, CA 94538, USA}
\author{Ian Crawley}
\altaffiliation{Present address: Nokia Bell Labs, Murray Hill, NJ 07974, USA}
\author{Ariel T. Sommer}
\email{ats317@lehigh.edu}
\affiliation{\lehigh}
\date{\today}
\begin{abstract}
Crossed optical dipole traps (ODTs) provide three-dimensional confinement of cold atoms and other optically trappable particles. 
However, the need to maintain the intersection of the two trapping beams poses strict requirements on alignment stability, and limits the ability to move the trap. 
Here we demonstrate a novel crossed ODT design that features inherent stability of the beam crossing, allowing the trap to move and remain aligned. 
The trap consists of a single high-power laser beam, imaged back onto itself at an angle to form a crossed trap. Self-aligning behavior results from employing an imaging system with positive magnification tuned precisely to unity.
We employ laser-cooled samples of $^6$Li atoms to demonstrate that the trap remains well-aligned over a \qty{4.3}{mm} travel range along an axis approximately perpendicular to the plane containing the crossed beams. 
Our scheme can be applied to bring an atomic cloud held in a crossed ODT close to a surface or field source for various applications in quantum simulation, sensing, and information processing.
\end{abstract}
\maketitle

\section{Introduction}
Advances in techniques for cooling and trapping atomic gases have enabled tremendous progress over the past few decades in quantum simulation, quantum sensing, and quantum information processing. Control over the position of an atomic sample often plays a crucial role in cold atom experiments. In quantum simulation, quantum gas microscopy relies on positioning atoms close to a high numerical aperture microscope objective~\cite{bakr2009quantum,miranda2012all-optical,cheuk2015quantum-gas}. Proximity of atoms to an RF antenna can be beneficial for
experiments employing strong RF magnetic fields to engineer atomic properties and observe novel phenomena~\cite{long2021spin,vivanco2023strongly}. Long-range transport  enables an atomic cloud to be moved to a different section of a vacuum system in complex setups~\cite{greiner2001magnetic,gustavson2001transport,gross2016all-optical,lee2020transporting,klostermann2022fast,seo2022moving-frame,bao2022fast,matthies2024long-distance}. 
In quantum sensing applications, sensitive detection of magnetic~\cite{behbood2013real-time, cohen2019cold} and electric~\cite{facon2016sensitive} fields from materials depends on the distance of atoms to the material surface~\cite{kruger2005cold, sorrentino2009quantum}. Quantum simulation, sensing, and information processing with hybrid atomic systems often rely on bringing an atomic sample to the vicinity of an optical device such as a high-finesse cavity~\cite{sauer2004cavity,ningyuan2016observation,kumar2023quantum-enabled} or hollow core fiber~\cite{peters2021loading,wang2022enhancing,song2024tightly}.

Methods to trap and position clouds of neutral atoms employ magnetic forces,
%~\cite{greiner2001magnetic,kruger2005cold}, 
optical forces, or a combination. All-optical trapping offers the advantages of faster evaporative cooling~\cite{barrett2001all-optical,kinoshita2005all-optical,fuchs2007molecular,clement2009all-optical,duarte2011all-optical,burchianti2014efficient,gross2016all-optical}, the ability to trap magnetically un-trappable states~\cite{hansen2011quantum}, freedom to tune the magnetic field independently~\cite{ohara2002observation}, and improved optical access~\cite{lee2020transporting,matthies2024long-distance}. In a single-beam optical dipole trap (ODT), dynamic trap positioning has been implemented in the transverse directions using a variety of methods~\cite{lengwenus2010coherent,endres2016atom-by-atom,barredo2016atom-by-atom,stuart2018single-atom,graham2022multi-qubit}, and in the axial direction by lens translation~\cite{gustavson2001transport,lee2020transporting,seo2022moving-frame} or with a focus-tunable lens~\cite{leonard2014optical,unnikrishnan2021long}. The axial confinement in an all-optical single-beam ODT is relatively weak, particularly when using larger beam diameters to achieve a large trap volume.
A crossed ODT solves this problem by forming a trap at the intersection of two beams that cross at an angle. The angle of the crossing controls the aspect ratio of the trap~\cite{grimm2000optical}, and the additional degrees of freedom of the second beam provide further control over the trapping potential~\cite{bourdel2004experimental,kinoshita2005all-optical,fuchs2007molecular,clement2009all-optical,anderson2019conductivity}. Crossed ODTs are therefore widely used for capturing atoms from a laser-cooled cloud. However, compared to the single-beam case, crossed traps are sensitive to misalignment, and tuning the position of a crossed ODT requires both beams to move in a coordinated way to keep them intersected. 
%While the independent adjustability of the two beams can be employed to actively maintain their alignment~\cite{frye2021bose-einstein}, 

%One strategy to keep a crossed ODT aligned during repositioning is to move the beams within a fixed plane so that they always intersect, as employed in \cite{anderson2019conductivity} to induce small displacements by modulating piezo-driven mirrors, and in \cite{tung2013ultracold,gross2016all-optical} to transport atoms over large distances by translating a lens.
%Another approach allows displacement of the plane containing the crossed beams via an acouso-optic deflector (AOD) by sending two separate beams through symmetric beam paths~\cite{chen2023dual-species}.

One method to keep a crossed ODT aligned during repositioning is to send two parallel beams through a shared lens that focuses the beams to the same point, forming a small-angle crossed trap at the focus~\cite{tung2013ultracold,gross2016all-optical}. Translation of the lens allows long-range transport of the trap along the axial direction. 
Another approach allows transverse displacement by splitting a beam after an acouso-optic deflector (AOD) and sending the two beams through symmetric paths~\cite{chen2023dual-species}.
However, to make maximal use of available laser power from a single light source, experiments often form a crossed ODT from a single recirculated beam that intersects with itself, giving twice the trap depth~\cite{fuchs2007molecular,williams2010universal,ottenstein2010few-body,mitra2018exploring,ottenstein2008collisional}.
The question then arises whether a recirculated crossed ODT can be implemented that remains aligned as its position is tuned.

In this paper, we show how to produce a recirculated crossed ODT that remains aligned while the trap is repositioned, and despite drift in the source beam. 
In particular, our method allows tuning the position of the trap over several millimeters along the axis perpendicular to the plane containing the crossed beams.  
We implement the self-aligning crossed ODT design in a new apparatus intended for studies of strongly interacting Fermi gases out of equilibrium~\cite{zhang2022transport}, to improve trap stability, and to facilitate tuning the trap position relative to a magnetic field saddle point.
To demonstrate the self-aligning property of the trap, we manually vary its position and employ a sample of $^6$Li atoms to show that the trapping beams maintain their intersection. 

In Section \ref{sec:Operating Principle} we describe the operating principle of our self-aligning crossed ODT. In Section \ref{secn:raytrace} we describe a numerical ray tracing analysis of the design. In Section \ref{sec:setup} we describe the experimental setup. In Section \ref{sec:results} we report experimental results demonstrating our setup with a gas of $^6$Li atoms. In Section \ref{sec:conclusion} we conclude.

\section{Operating principle of the self-aligning trap}\label{sec:Operating Principle}
\subsection{General Case of a Recirculated Crossed ODT}
\begin{figure}\includegraphics[width=3.5in]{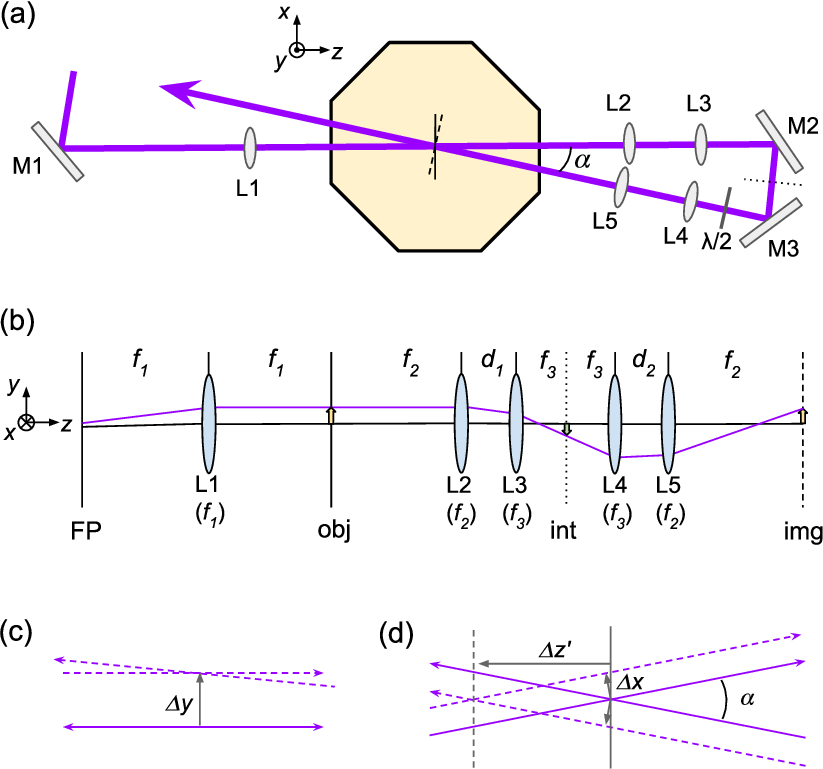}
	\caption{\label{fig:setup}(a) Schematic of the self-aligning recirculated crossed ODT (top-down view) showing the trap beam passing twice through the vacuum chamber (octagon). The short solid line in the chamber represents the horizontal axis in the object plane ($x$), while the dashed line represents the horizontal axis of the image plane ($\bar{x}$). The two planes share a common $y$ axis (out of the page). The dotted line between M2 and M3 is the intermediate focal plane. (b) Unfolded lens system showing the Fourier plane (FP), object plane (obj), intermediate focal plane (int) and image plane (img). The latter three are indicated in solid, dotted, and dashed lines, respectively, to match the schematic in (a). The purple line shows the path of the trapping beam after a vertical tilt of M1.  (c) Vertical adjustment of M1 moves the first and second pass foci by the same amount $\Delta y$ when the vertical magnification is tuned to 1. (d) Horizontal adjustment of M1 causes the two trapping beams to move away from one another symmetrically. The beams continue to intersect if they lie in the horizontal plane, which happens for $y=0$, and for non-zero $y$ in a setup where $d_1+d_2=2(f_2+f_3)$.  The crossing point moves along the longitudinal ($z'$) axis of the trap. }
\end{figure}

We consider a crossed ODT formed in the horizontal ($xz$) plane as shown in Figure \ref{fig:setup}. A single beam passes twice through a vacuum chamber, intersecting itself at an angle $\alpha$ to form a crossed trap. Both passes of the beam are focused at the crossing point, where atoms are trapped.

\begin{figure*}[ht]
    \centering
    \includegraphics[width=6.5in]{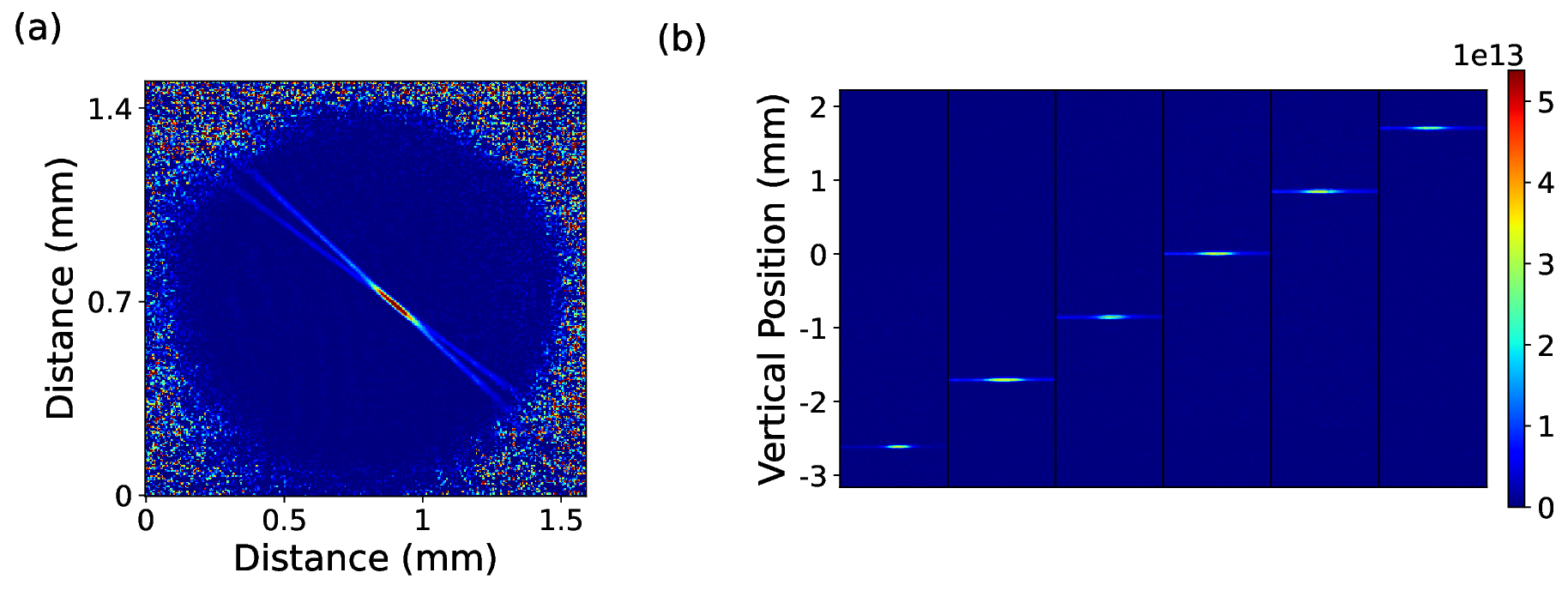}
    \caption{(a) Top-down view of the crossed ODT. (b) Side view of the crossed ODT after moving the trap from $y=0$ to different vertical positions. The second-pass focus is imaged precisely back onto the first-pass focus, leaving the trap intact and able to hold atoms. The color scale in (a) and (b) indicates the two-dimensional column density in state $\ket{2}$ atoms/m$^2$, obtained from absorption imaging.}
    \label{fig:ODTpictures}
\end{figure*}

\begin{figure}
    \centering
    \includegraphics[width=3.15in]{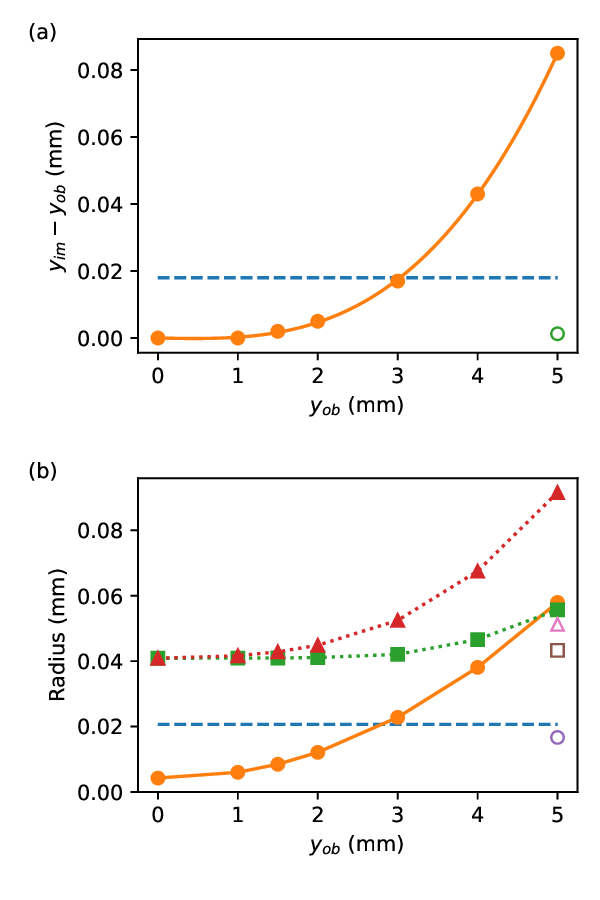}
    \caption{Numerical simulation. (a) Misalignment $y_\text{im}-y_\text{ob}$ due to geometric aberration. As in the main text, $y_\text{ob}$  is the vertical position of the reference ray (ie, the beam center) in the object plane, and  $y_\text{im}$ is the position of the reference ray when it reaches the image plane. Solid circles: numerical simulation using the parameters of our experimental setup. 
    The curve is from a fit to Eqn. (\ref{eqn:cubic}). The dashed line indicates the value $0.45 w_0 = \SI{18}{\micro\meter}$, for reference, as discussed in text. Open circle: $y_\text{im}-y_\text{ob}$ for a setup where $C=0$, showing that even at the most extreme end of the range ($y_\text{ob}=\qty{5}{mm}$), the misalignment is negligible. (b) Resolution and beam sizes. Solid circles: geometric RMS radius in image plane. Dashed line: Airy disk radius. Solid squares and solid triangles: beam size as $1/e^2$ semi-diameters in the image plane along $\bar{x}$ and $\bar{y}$ axes, respectively, from diffraction integral. Open symbols: same quantities as corresponding solid shapes, but for $C=0$ setup.}
    \label{fig:OSLOresults}
\end{figure}

The focus of the second-pass beam is formed using a lens system that images the first-pass focal plane (object plane) onto the second-pass focal plane (image plane). A paraxial ray in the object plane is mapped onto a ray in the image plane by a formal $2 \times 2$ ray matrix:
\begin{equation}\label{eqn:raymatrix}
\begin{pmatrix} \bm{\bar{\rho}}\\d\bm{\bar{\rho}}/d\bar{z}\end{pmatrix}_{\text{im}}
=
\begin{pmatrix} A & B \\ C & D\end{pmatrix}
\begin{pmatrix} \bm{{\rho}}\\d\bm{{\rho}}/d{z}\end{pmatrix}_{\text{ob}}
\end{equation}
Here $\bm{\rho}=(x,y)$ are transverse coordinates, and $z$ is the longitudinal coordinate, in the object coordinate system. The barred variables ($\bar{x},\bar{y},\bar{z}$) refer to the image coordinate system. The subscripts ``ob'' and ``im'' refer to object and image points, respectively. The image plane is defined to minimize the beam spot size including aberrations, but is close to the paraxial image plane where $B=0$. %The matrix element $A$ gives the magnification of the imaging system between the two local coordinate systems. 
Due to the folding of the optical path by mirrors M2 and M3, the object and image planes intersect at an angle of $\pi-\alpha$. We can relate the coordinates of a point P in the two coordinate systems by a rotation matrix:
\begin{equation}
\begin{pmatrix} \bar{x}\\ \bar{y} \\ \bar{z} \end{pmatrix}_\text{P} =
\begin{pmatrix} -\cos\alpha &0 & -\sin\alpha \\
0 & 1 & 0\\
\sin\alpha & 0 & -\cos\alpha\end{pmatrix}
\begin{pmatrix}x\\y\\z\end{pmatrix}_\text{P}
\end{equation}
For small $\alpha$, then, $x_\text{im} \approx -\bar{x}_\text{im} = -A x_\text{ob}$ so that, in a fixed coordinate system, the horizontal magnification is approximately $-A$. Meanwhile, for any $\alpha$, $y_\text{im}=\bar{y}_\text{im} = A y_\text{ob}$, so the vertical magnification is $A$ in both global and local coordinates.

A conventional method of forming a recirculated crossed ODT employs a  pair of lenses to refocus the beam~\cite{fuchs2007molecular,williams2010universal,ottenstein2010few-body,mitra2018exploring} and gives $A<0$. In a typical setup, the lenses have equal focal lengths and are each located one focal length from the atoms, so $A\approx -1$. The vertical magnification is then negative, so that a vertical movement of the first-pass focus causes an opposite movement of the second-pass focus. 
%Attempting to reposition the trap by moving the first-pass beam in the vertical direction then causes the trap to lose alignment. 
The alignment of the trap is therefore sensitive to drift in the pointing of the first-pass beam. Furthermore, any attempt to reposition the trap along the axis perpendicular to the crossed beams (i.e., vertically) causes the trap to lose alignment.

\subsection{Self-aligning Configuration}

In this work, we study an alternative setup that allows repositioning of the trap without loss of alignment. In our setup, shown in Fig. \ref{fig:setup}(a) and (b), the imaging system that refocuses the second-pass beam consists of two sub-systems, each with a negative vertical magnification, resulting in a net positive vertical magnification $A>0$ that we tune near to 1.  The first sub-system consists of lenses L2 and L3, with focal lengths $f_2$ and $f_3$, respectively, separated by a distance $d_1$, and refocuses the beam onto an intermediate focal plane with a magnification of $-f_3/f_2$. The second sub-system consists of lenses L4 and L5, with focal lengths $f_3$ and $f_2$, respectively, separated by a distance $d_2$, and images the intermediate focal plane onto the final image plane in the vacuum chamber, with a magnification of $-f_2/f_3$. The net magnification is then 1. The ray matrix appearing in (\ref{eqn:raymatrix}) becomes approximately
\begin{equation}\label{eqn:ourmatrix}
\begin{pmatrix}
A & B \\ C & D 
\end{pmatrix} = \begin{pmatrix} 1 & 0\\
\displaystyle
\frac{2(f_2+f_3)-(d_{1}+d_{2})}{f_2^2} & 1\end{pmatrix}
\end{equation}
In practice, we fine-tune the positions of the lenses to achieve a magnification of unity, as described later.

To facilitate repositioning the trap, we place a mirror (M1) in the Fourier plane of the trap before lens L1. Tilting M1 then displaces the beam after L1 without changing its direction, as shown in Figure \ref{fig:setup}(c), allowing the beam to be moved over a larger range without leaving the aperture of the downstream optics. In principle, M1 can be a galvo-driven mirror~\cite{lengwenus2010coherent}, or replaced with an acousto-optic deflector~\cite{endres2016atom-by-atom,barredo2016atom-by-atom,chen2023dual-species}, to allow real-time transport of the cloud of atoms in the trap. For this demonstration, we adjust M1 manually and show that the trap remains self-aligned, as illustrated in Fig.~\ref{fig:ODTpictures}.

Unlike the vertical behavior, horizontal displacement of the first-pass beam with $A=1$ causes opposite movement of the second-pass beam in the vacuum chamber. Under a pure horizontal displacement, the beams stay in the same plane and remain intersected. The crossed trap then moves along its longitudinal axis ($z'$), as illustrated in Fig \ref{fig:setup}(d), by a displacement $\Delta z' = \Delta x/\sin(\alpha/2)$. Here $\Delta x$ is the displacement of the first-pass beam focus in the object coordinate system. The primed coordinates refer to the principle axes of the trap, and are related to the object coordinates by a rotation of $\pi-\alpha/2$ about $y$. For a situation requiring large displacements in both vertical and horizontal directions, one can set $C= 2(f_2+f_3)-(d_1+d_2)=0$. However, in our case, we prefer a shorter path length for a simpler setup and choose $d_1$ and $d_2$ to be smaller than $f_2+f_3$.

One might ask whether it is necessary to introduce the intermediate image plane in order to achieve a positive magnification ($A>0$). Given the mirror arrangement shown in Figure \ref{fig:setup}(a), no choice of lenses can produce a positive magnification without introducing an intermediate focus, as shown in Appendix \ref{app:int}. 

In our implementation, the focal lengths of the lenses shown in Figure \ref{fig:setup} are: $f_1= 250$ mm, $f_2 = 350$ mm, and $f_3 = 150$ mm. Lenses L1--L5 are plano-convex spherical singlets of 25.4 mm diameter. The distance between L2 and L3 is $d_1=$ 133 mm and between L4 and L5 is $d_2=$ 127 mm. The crossing angle is $\alpha =$ 3.5$^\circ$. A collimated beam of about \qty{2}{mm} $1/e^2$ intensity radius enters the system at M1.

\section{Numerical Ray Trace Analysis}
\label{secn:raytrace}
Numerical ray tracing gives the predicted tuning range of the trap in the vertical direction. 
We employ the OSLO software program to assess the effects of finite lens aperture and geometric aberrations, which are not included in the paraxial model described above.
The simulation models a gaussian beam using a gaussian distribution of rays of which the central (reference) ray passes through the object plane at $y_\text{ob}$ and slope zero, similar to Fig.~\ref{fig:setup}(b). Additional rays fan out from that point to produce a \SI{3}{\milli\meter} $1/e^2$ radius gaussian spot on the first surface of L2, corresponding to an effective gaussian beam radius of $w_0=\SI{40}{\micro\meter}$ in the object plane. The size of the focus in the actual experimental setup is similar, as detailed in Section \ref{sec:setup}. The lens positions and properties are chosen to match the experimental setup. We fine-tune the lens positions to achieve a magnification of 1, as judged by tracing a ray with $y_\text{ob}=\SI{1}{\milli\meter}$. We then vary $y_\text{ob}$ to model the effect of tuning the position of the first-pass beam.

The system performs well in simulation up to about $y_\text{ob}=\SI{3}{\milli\meter}$, at which point several effects begin to limit the performance. The finite aperture of the lenses imposes one limitation. 
We aim to keep the beam center at least two $1/e^2$ radii away from the lens apertures, which we take to be 11 mm in radius, to avoid clipping the beam. This necessarily limits $y_\text{ob}$ to be less than \SI{5}{\milli\meter} to avoid clipping at L2. However,
at $y_\text{ob}=\SI{3}{\milli\meter}$, the beam gets close to the edge of L4, with the 
rays at twice $1/e^2$ attaining a $y$ coordinate of $\SI{-10.5}{\milli\meter}$. This larger deviation of the rays from the optical axis near L4 can be avoided by using $C=0$, as described later.

Aberrations cause the position $y_\text{im}$ of the reference ray in the image plane (ie, the second-pass focus) to deviate from the center $y_\text{ob}$ of the first-pass beam. Figure \ref{fig:OSLOresults}(a) shows the deviation $y_\text{im}-y_\text{ob}$ versus trap position $y_\text{ob}$. By symmetry, $y_\text{im}$ is an odd function of $y_\text{ob}$, so to 3rd order,
\begin{equation}\label{eqn:cubic}
    y_\text{im}= Ay_\text{ob} + \sigma y_\text{ob}^3
\end{equation}
The constant $\sigma$ contains contributions from all five third-order Seidel coefficients. Fitting to the simulation results gives $\sigma = \SI{7e-4}{\per\square\milli\meter}$. The tuning procedure based on setting $y_\text{im}=y_\text{ob}$ at $y_\text{ob}=\SI{1}{\milli\meter}$ technically results in $A\approx 1-\sigma\cdot(\SI{1}{\milli\meter})^2$, however, the difference is negligible.

We aim to keep $|y_\text{im}-y_\text{ob}|$ less than $0.45 w_0$. 
In the absence of additional broadening of the second-pass focus, this amount of offset of the beams would result in a 10\% reduction in trap depth for the crossed trap, a 15\% reduction of the vertical trap frequency, a 5\% reduction in the two horizontal trap frequencies.
At $y_\text{ob}=3$ mm, we find $y_\text{im}-y_\text{ob} = \SI{17}{\micro\meter}$, or $0.43 w_0$, close to the desired limit.

Geometric aberration due to large $y_\text{ob}$ can also degrade the trap by broadening the beam focus in the image plane. Figure \ref{fig:OSLOresults}(b) shows the root-mean-square (RMS) spot size from geometric ray tracing. 
For $y_\text{ob}=0$, ray tracing predicts a diffraction-limited spot with a geometric RMS radius of about \SI{4}{\micro\meter}. 
The geometric RMS spot size increases with increasing $y_\text{ob}$,  and exceeds the Airy disk radius for $y_\text{ob}$ greater than about \SI{3}{\milli\meter}. Figure \ref{fig:OSLOresults}(b) also shows that the total beam size at the focus, including both diffraction and geometric aberration, begins to increase noticeably around $y_\text{ob}=\SI{3}{\milli\meter}$. Due to the symmetry of the system, the position can also be tuned in the opposite direction down to $y_\text{ob}=\SI{-3}{\milli\meter}$.
In summary, ray tracing indicates that the trap in our setup can be tuned by $\pm\qty{3}{mm}$, for a \qty{6}{mm} travel range, while staying self-aligned and well focused.

We also investigated a hypothetical setup in which the distances $d_1$ and $d_2$ are chosen to make $C=0$. The results, shown as the open points in Fig.~\ref{fig:OSLOresults}, indicate that the system performs well throughout the full range of $\pm \SI{5}{\milli\meter}$. At that point, the system is limited by the aperture of the lenses, and is also close to being no longer diffraction limited. Remarkably, the position error $y_\text{im}-y_\text{ob}$ remains negligible, at about \SI{1}{\micro\meter}. 

\section{Experimental Setup}
\label{sec:setup}

Light for the optical dipole trap comes from a linearly polarized \SI{200}{\watt} IPG Photonics fiber laser at \SI{1064}{\nano\meter} wavelength, which we operate at 70\% power. The power is stabilized and controlled using an acousto-optic modulator (AOM; Gooch \& Housego AOMO 3110-197).
Transmission through a polarizing beamsplitter plate before M1 ensures pure horizontal polarization. 
A halfwave plate on the second pass rotates the polarization to the vertical to prevent interference with the first-pass beam. 
Light transmitted through a backside-polished mirror (M2) is sent to a photodiode for stabilization. 
The first- and second-pass beams have respective $1/e^2$ intensity waist radii of $(w_{1x}, w_{1y})= (\qty{47(1)}{\micro m}, \qty{33(1)}{\micro m})$ and $(w_{2\bar{x}}, w_{2\bar{y}})= (\qty{51(4)}{\micro m}, \qty{41(4)}{\micro m})$,  and maximum optical powers of \qty{78}{W} and \qty{74}{W}. For these parameters, gravitational sag of lithium atoms is expected to be negligible ($<\qty{1}{nm}$).

%Note about the waists:
%We measured the beam sizes with the laser and AOM at full standard operating power, to capture thermal lensing in the AOM. The beams were measured on a camera after a beam sampler and optical attenuators.
%first pass measured on 3.26.2024
%second pass measured on 5.2.2024 and 3.26.2024
%available measurements were averaged
%error estimates are 1 std for the second pass
%for the first pass, it is based on fit uncertainty
%  
%Optical powers measured 5.1.2024; assuming ~1% reflection from AR coated window

We test our setup by loading $^6$Li atoms into the crossed ODT from a laser cooled cloud. 
First, a magneto-optical trap (MOT) on the D2 line collects about $5\times 10^8$ atoms in \SI{5}{\second} from a Zeeman-slowed atomic beam. The ODT light is applied to the cloud during the final  \SI{2}{\second} of MOT loading, to minimize effects of thermal lensing by allowing the optics to reach a steady temperature. The MOT employs three retro-reflected beams of red-detuned cooling and repumping light, with powers of 12 mW and 3.1 mW per beam, respectively, and beam diameters of about 10 mm. The MOT quadrupole magnetic field produces a \qty{32}{G/cm} gradient in the axial direction. Small bias fields ($<$\qty{10}{G}) control the MOT position. The MOT has a temperature of about \qty{2}{mK}. To reduce the temperature to \qty{500}{\micro K} and increase the density, we apply a compressed MOT (cMOT) phase for \qty{5}{ms} by reducing the cooling and repumping light detunings from \qty{-35}{MHz} and \qty{-27}{MHz}, respectively, to \qty{-10}{MHz} each, while reducing their intensity to 1-2\% of their initial values. The magnetic field gradient is kept constant.  The Zeeman slower coils are shut off after the first \qty{3}{ms} of the cMOT phase, together with the bias field along the Zeeman slower axis. At the end of the cMOT phase, we switch off the remaining bias fields and the MOT quadrupole magnetic field. The cooling and repumping light remain on for a \qty{0.1}{ms} dwell time before proceeding with gray molasses cooling.

Gray molasses cooling on the D1 line~\cite{burchianti2014efficient}  is applied for 1.5 ms after the compressed MOT dwell time, and brings the temperature down to about \qty{150}{\micro K}. To deliver D1 light to the apparatus, we combine four beams, the D1 and D2 cooling and repumping beams, on a $4\times 4$ fiber splitter array. Three output fibers are used for the three axes of cooling, and the fourth output is used as a monitor. The D1 cooling and repumping beams have 13 mW and 1.1 mW of power per beam, and are blue-detuned by \qty{25}{MHz} from the $F=3/2$ to $F'=3/2$ and $F=1/2$ to $F'=3/2$ transitions, respectively. 

After extinguishing the gray molasses light, we optically pump atoms into the $F=1/2$ ground state manifold. Atoms not held in the crossed ODT now fly away freely. We ramp up a magnetic field in the vertical direction to \qty{286}{G} over 15 ms and hold the cloud for \qty{500}{ms} to allow equilibration. Atoms then occupy the two lowest hyperfine states, $\ket{1}$ and $\ket{2}$. We primarily image state $\ket{2}$, via absorption of a \SI{10}{\micro\second} pulse of resonant light. To image state $\ket{1}$, we instead ramp to \qty{338}{G}, using the same light frequency. For most purposes, we   image along an axis in the horizontal plane, at \qty{45}{\degree} to the first-pass ODT beam, using horizontal linear polarization. A second imaging system operates in the vertical direction to visualize the beam crossing as in Fig.~\ref{fig:ODTpictures}(a).

\section{{Testing the self-aligning trap with $^6\text{Li}$ atoms}}
\label{sec:results}

\begin{figure}
\centering
	\includegraphics[width=3in]{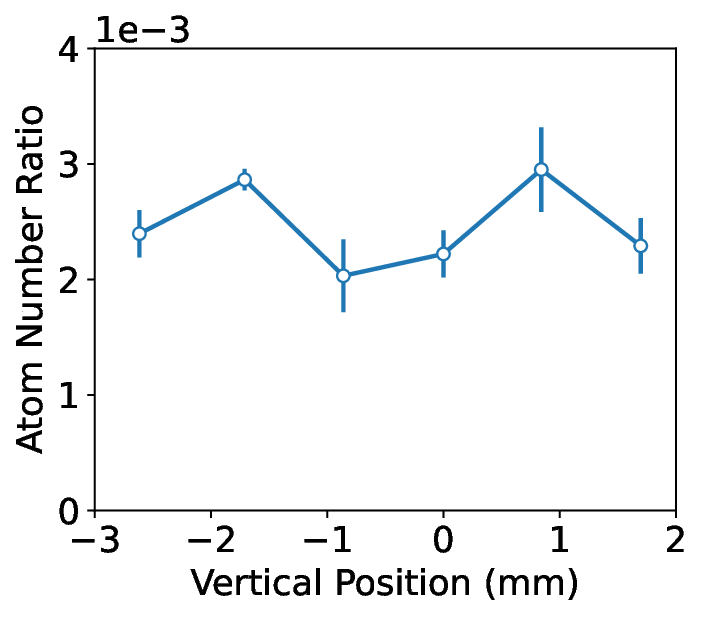}
 	\caption{\label{fig:AtomLoading} Fraction of atoms transferred from gray molasses to the crossed ODT vs position of the ODT. The gray molasses is also moved to match the ODT location by adjusting the vertical bias field.}
\end{figure}

\subsection{Tuning to Unity Magnification}
Before testing the self-aligning performance of the trap, we tuned the positions of lenses L4 and L5 through an iterative procedure to bring the magnification to 1.
We first aligned the two foci of the crossed trap, with the help of images of the trapped atoms.
We then moved the first-pass focus in the vertical ($y$) direction by about a millimeter and loaded atoms into the ODT. At first, the magnification was sufficiently far from 1 that the beams no longer overlapped afterwards, and we could see two separate atomic clouds. From images of the two clouds, we determined the positions $y_\text{ob}$ and $y_\text{im}$ of the first- and second-pass foci, respectively, giving the magnification $M\equiv y_\text{im}/y_\text{ob}$ at that step of iteration. 
To improve the magnification, we adjusted the positions $z_4$ and $z_5$ of lenses L4 and L5 based on a ray optics calculation. Theoretically, the derivative $dA/dz_4$ of the paraxial magnification with respect to the position of lens L4, and subject to the constraint that lens L5 is simultaneously adjusted to keep the image plane fixed, is  $dA/dz_4 = \qty{-0.017}{mm^{-1}}$; meanwhile, $dz_5/dz_4=4.4$. Using these derivatives, we estimated the displacements $dz_4$ and $dz_5$ needed to bring the magnification to 1 without shifting the focus. After applying the adjustments, and fine-tuning L5 using the atoms to correct any remaining offset in the axial position of the focus, we measured the magnification again, then adjusted L4 and L5 a second, and final, time.

\begin{figure}
    \centering
    \includegraphics[width=3in]{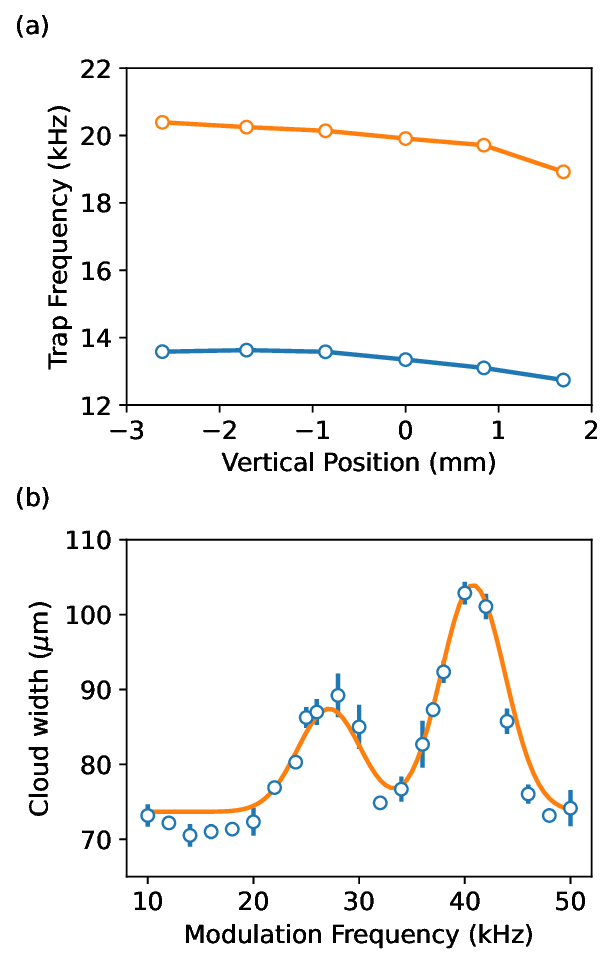}
    \caption{(a) Transverse trap frequencies as a function of ODT position. 
    (b) Typical modulation spectrum, showing cloud size (one standard deviation of the vertical density distribution) after 0.1 ms time-of-flight. The peaks correspond to 2 times the transverse trap frequencies.}
    \label{fig:freq}
\end{figure}

After the second update of L4 and L5, 
moving the trap from its central, aligned position to a new position $y_\text{ob}=\qty{1.7}{mm}$ produced no visible splitting of the atomic density distribution versus $y$, indicating that the two foci moved by the same amount to within about a beam radius. To measure the magnification at that point, we applied a new procedure to detect the misalignment $y_\text{im} - y_\text{ob}$ with a precision finer than the atomic cloud size. 
We measure the small misalignment by 
relaying the second-pass focus onto an auxiliary camera, using a sample of the trap light transmitted by a backside-polished mirror. We measure the positions $\tilde{y}_1$ and $\tilde{y}_2$ of the beam center on the auxiliary camera before and after moving the first-pass beam to $y_\text{ob}$. We then intentionally misalign the second-pass beam in the vertical direction using M3, enough to create two separate optical traps. We measure the vertical displacement $dy$ between the two atomic cloud centers versus the position $\tilde{y}$ of the beam on the auxiliary camera as we scan M3.   A linear fit then gives the value  $\tilde{y}_3$ at which $dy=0$. The value $\tilde{y}_3$ identifies the position of the first-pass focus in the coordinate system of the auxiliary camera. Therefore, $y_\text{ob} \propto \tilde{y}_3 - \tilde{y}_1$, and $y_\text{im} \propto \tilde{y}_2 - \tilde{y}_1$. Finally, the magnification of the self-aligning crossed ODT is given by $M=(\tilde{y}_2-\tilde{y}_1)/(\tilde{y}_3-\tilde{y}_1)$.  In the final configuration, the magnification was indistinguishable from 1. Using $y_\text{ob} = -$1.7 mm gave $M=0.995$ and using $y_\text{ob} = $1.7 mm gave $M=1.004$, which we summarize as $M=1.000(5)$.

\subsection{Evaluation of Self-Aligning Performance}
To evaluate the performance of the self-aligning crossed ODT, the trap is tuned over a range of positions along the vertical direction using mirror M1. At each trap position, we adjust the vertical MOT bias field to center the laser-cooled cloud on the crossed ODT and maximize the number of atoms transferred. The trap is evaluated by measuring the loading efficiency and the transverse trapping frequency at each position. 

The location of the crossed ODT was adjusted over a vertical range of 4.3 mm. Figure~\ref{fig:ODTpictures}(b) shows images of the trapped cloud at each of six positions tested, demonstrating qualitatively that the trap remains aligned. To determine the loading efficiency, the atom populations of the gray molasses cloud and the crossed ODT were measured at each position. The loading efficiency is shown in Fig. \ref{fig:AtomLoading}. On average, we obtain $3\times 10^8$ atoms in the gray molasses cloud and $7.5\times 10^5$ atoms in the crossed ODT, for an average loading efficiency of $2.5\times 10^{-3}$. For vertical positions beyond $-2$ mm, the atom population in both traps decreases, due to reduction in the MOT atom number. In the other direction, the travel range is limited by the lens apertures due to imperfect lens centering. However, the loading efficiency stays relatively constant throughout the range measured, demonstrating the self-aligning property of the crossed ODT.

The trap frequencies serve as another indicator of alignment.
Transverse trap frequencies were measured at each ODT position by parametric heating at a magnetic field of \qty{286}{G}, and are shown in Fig. \ref{fig:freq} (a).
The ODT optical power was modulated sinusoidally 
by applying 800 cycles of amplitude modulation to the AOM RF drive. The average optical power in the first-pass beam was reduced to \qty{68}{W} for better linearity, with a modulation amplitude of \qty{1.1}{W}.
At this magnetic field, temperature, and density, the gas is in the nearly collisionless, classical regime in a nearly harmonic potential, where its collective mode frequencies are integer multiples of the trap frequencies $\nu_i$~\cite{guery-odelin1999collective,stringari1996collective,heiselberg2004collective}. Parametric resonances are expected at $2\nu_i$ and sub-harmonics~\cite{balik2009imaging-based}.

Parametric heating of the cloud is detected by measuring the cloud size after \qty{0.1}{ms} time-of-flight, which gives greater sensitivity than measuring atom loss~\cite{balik2009imaging-based}. A typical modulation spectrum is shown in Fig. \ref{fig:freq}(b). The resonances correspond to $2\nu_i$ for the two transverse trap frequencies. 
Due to the ellipticity of the trapping beams, the transverse trap frequencies are several kHz apart. 
We extract the peak positions by fitting the spectra to a sum of two gaussians. 
The highest trap frequencies obtained for the two axes are $\nu_{x} = \qty{13.6}{kHz}$ and $\nu_{y} = \qty{20.4}{kHz}$, and decrease by 
about 7\% each
across the range of trap positions. In comparison, the amount of vertical misalignment that would lead to a 10\% reduction in trap depth, including the beam ellipticities, would cause reductions by 7\% and 13\%, respectively, in the trap frequencies.  The trap frequencies stay above this threshold, indicating that the crossed ODT remains well aligned to within the desired specifications.

As a final point, we note that the self-aligning configuration greatly improved the trap stability in daily use compared to a conventional configuration with negative vertical magnification. This suggests that the dominant source of alignment drift in this setup comes from the first-pass beam, possibly due to strong thermal lensing in the AOM~\cite{simonelli2019realization}. We therefore find this setup advantageous even when the trap does not need to be repositioned dynamically.

\section{Conclusion}
\label{sec:conclusion}
We demonstrated a self-aligning crossed optical dipole trap using a recirculated trapping beam in an approximately horizontal plane.
The presence of an intermediate focus gives a positive lateral magnification of the overlapping foci. Tuning the vertical magnification to 1.000(5) yielded a self-aligning configuration, in which the foci remain overlapped as the first focus is moved in the vertical direction. 
Numerical ray tracing predicts that this scheme works well even in the presence of geometric aberration. We confirmed these predictions by loading $^6$Li atoms into the trap throughout a \qty{4.3}{mm} range of trap positions and observing robust atom loading and trap frequencies without requiring any realignment to account for the position change. 

The self-aligning property increases the stability of the recirculated crossed ODT and provides the ability to transport the trapped atoms if required.
Our numerical ray tracing analysis predicts that the travel range of the trap can be extended further with a slight variation on the setup (using $C=0$), and by using optical components with larger apertures.
In the context of dynamic positioning, we envision this method being useful for bringing atoms in a recirculated crossed ODT close to a surface. Lateral movement of the trap position would allow bringing the sample to
within a few times the beam waist, typically tens of microns, from a surface without subjecting the surface to the trapping beams. This could be used in field sensing from surfaces with cold atom sensors, or to evaporatively cool a cloud of atoms to quantum degeneracy before transporting it to the vicinity of a hollow core fiber for quantum optics applications.

\begin{acknowledgements}
We would like to thank several current and former undergraduate students who contributed to constructing components of the apparatus used in this work: Eason Shen, Amondo Lemmon, George Awad, and REU students Jianyi Chen, Cameron Brady,  and Mary Kate Pasha (NSF REU Site Award No. 1852010).
This work was supported by the National Science Foundation (Award No. 2110483) and a Lehigh University Faculty Innovation Grant (FIGAWD274). M.M-M. acknowledges support from a Lee Fellowship through the Lehigh University Physics Department.
\end{acknowledgements}

\appendix
\section{Necessity of an intermediate focus}\label{app:int}
Our optical system achieves a positive magnification by producing an intermediate focus partway through the imaging system. Is it possible to achieve a positive magnification without an intermediate image plane? Here we show that the intermediate focal plane is required in a system where the optical axis lies within a single plane (the horizontal plane in our system).

We consider the vertical component of the paraxial rays. The ray transfer matrix relates the input and output rays of the optical system:
\begin{equation}
\begin{pmatrix}
y_f \\ \theta_f
\end{pmatrix} = \begin{pmatrix}A & B \\ C & D\end{pmatrix}\begin{pmatrix}y_i \\ \theta_i\end{pmatrix}
\end{equation}
The condition $B=0$ ensures that the system images rays from a given point in the input (object) plane to a single point in the output (image) plane regardless of their angle. A classical optical system (composed, for example, of lenses, mirrors, and free-space propagation) has the property~\cite{guenther1990modern}:
\begin{equation}
AD-BC = n/n'
\end{equation}
where $n$ and $n'$ are the indices of refraction of the medium in the input and output planes, respectively. This condition expresses conservation of phase-space volume and relates to the concept of the optical invariant. In our case, $n'=n$, $B=0$, $A=M$ (lateral magnification) and $D=M_\gamma$ (angular magnification), giving:
\begin{equation}
M M_\gamma = 1
\end{equation}
A ray starting at $y_i=0$ and arbitrary positive angle $\theta_i>0$ %(for example, a marginal ray) 
will enter the lens system with positive position $y_1 > 0$. By definition, this ray will cross the optical axis at any image plane, because it started at $y_i=0$. If no intermediate image plane occurs, this ray never crosses the optical axis within the lens system, and must exit with $y > 0$. Therefore, approaching the final focus, this ray will have a negative final angle $\theta_f < 0$. The angular magnification $M_\gamma = \theta_f/\theta_i$ is therefore negative, and the lateral magnification $M$ must also be negative. Therefore, a lens system with no internal image planes, and a planar optical axis, will have a negative magnification in the direction perpendicular to the plane containing the optical axis.

To achieve a positive magnification (relative to a fixed axis) without introducing an intermediate image plane, one could employ a non-planar optical path. In particular, one can leverage the inversion of the axis that lies within the plane of incidence upon reflection from a mirror, as in a telescope star diagonal.

\bibliography{Self-aligned_cODT}
\end{document}